\documentclass[10pt,a4paper]{article}
\usepackage[affil-it]{authblk}
\usepackage{graphicx}
\usepackage{subfigure}
\usepackage{cite}
\usepackage{textcomp}
\usepackage{enumitem}
\usepackage{color}
\begin{document}
\title{Conductivity and Discharge Guiding Properties of Mid IR Laser Filaments}

\author{D. Mongin, E. Schubert, J. Kasparian, J.P. Wolf}
\affil{GAP-Biophotonics, University of Geneva 22, ch. de Pinchat, 1211 Geneva 4, Switzerland}

\author{V. Shumakova, S. Alisauskas, A. Pugzlys, A. Baltuska}
\affil{Photonics Institute, Vienna University of Technology, Gußhausstraße 27/387, 1040 Vienna, Austria}

\maketitle
\begin{abstract}
The electric properties of mid-IR filaments in air have been investigated in comparison to their more traditional near IR counterparts. Although the number of ions left after the pulse is similar for both spectral regions, filaments at 3.9~{\textmu}m produce lower electron densities and lower pressure depression, which prevent them in the present conditions (25~mJ,  90~fs) to trigger or guide HV discharges (130~kV over 12~cm). We conclude that mid-IR filaments require significantly higher energy/power levels than their near IR counterparts for applications related to lightning control and for fully taking advantage of their unique propagation properties (single, large diameter filaments over long distances). 
\end{abstract}
\section{Introduction}
Femtosecond mid infrared (2-10~{\textmu}m) filaments have recently attracted much attention because of their unique properties as compared to their more traditional counterparts in the visible and near infrared spectral range. More precisely, mid-IR filaments appear insensitive to spatial break-up into multiple filaments, and are expected to produce single millimeter sized “optical pillars” bearing very high intensities over hundreds of meters \cite{mitrofanov_mid-infrared_2015,panagiotopoulos_super_2015,geints_single_2014}. The difference in the filamentation process between the near-IR and the mid-IR regimes relates, to a large extent, to the $\lambda^2$ dependence of the critical power: while P$_c$ $\approx $ 3.3~GW \cite{couairon_femtosecond_2007} at 800~nm  and 5.3~GW \cite{houard_study_2016}  at 1030~nm, P$_c$ reaches 150~GW at 3.9~{\textmu}m \cite{mitrofanov_mid-infrared_2015}. The first experimental observation of mid-IR filaments in gases was achieved in high pressure Argon \cite{kartashov_white_2012}  using advanced optical parametric chirped pulse amplification (OPCPA) technologies at 3.9~{\textmu}m \cite{andriukaitis_220-fs_2015}. Thanks to further development of this unique system, reaching now the sub-TW level, mid-IR filaments could be produced in atmospheric air for the first time \cite{mitrofanov_mid-infrared_2015}. Significant work is currently carried out for developing mid-IR multi-TW sources, including scaling up OPCPAs and ultrashort laser seeded high pressure CO2 amplifiers \cite{polyanskiy_chirped-pulse_2015,haberberger_fifteen_2010,Lassonde_high_2015}.
The atmospheric applications of such intense and long-distance spanning single filaments are numerous, ranging from remote sensing of multiple pollutants to lightning control. For the latter application, investigating the electrical properties of these mid-IR waveguides is essential. In this letter, we present the first measurements of plasma density, electric conductivity, HV-discharge guiding properties and neutralization capability \cite{schubert_remote_2015}  of intense mid-IR filaments in air, and compare the results to near IR wavelengths, namely 800~nm and 1030~nm.

\section{Experimental setup}

\begin{figure}[h]
\centering\includegraphics[trim = 0 0 0cm 11cm,clip,width=0.9\textwidth]{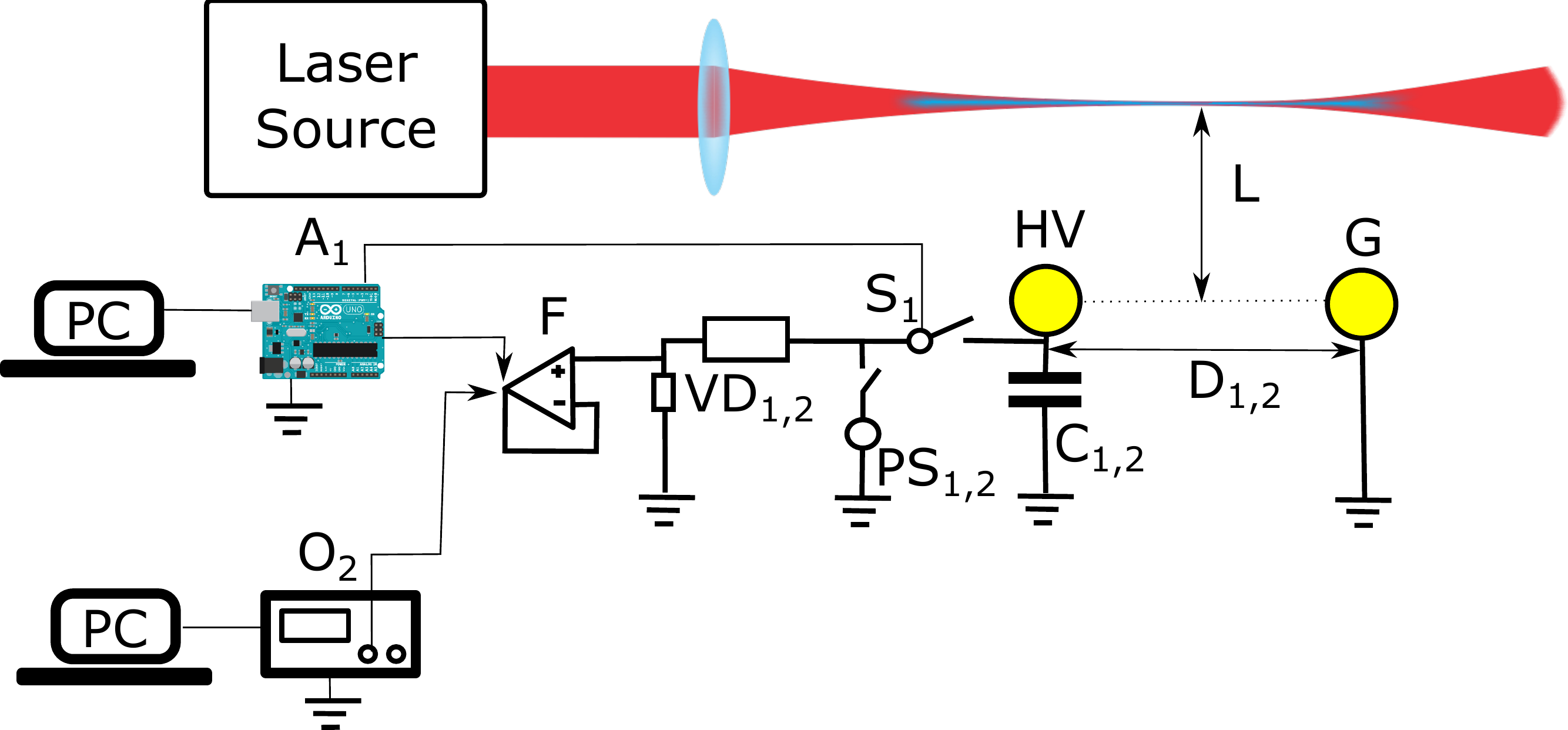}
\caption{Experimental setup for investigation of conductivity (1) and discharge guiding (2) properties of NIR and MIR filaments. HV –- high voltage electrode, G -– grounded electrode, PC –- computer, A$_1$ -– Arduino acquisition and control of switch S$_1$, O -– oscilloscope acquisition, F -- follower circuit, VD$_1$ and VD$_2$ -– voltage divider correspondingly 1:4000 divider of 300 G$\Omega$ impedance and 1:20000 divider of 400 G$\Omega$ impedance, PS$_1$ and PS$_2$ -– correspondingly 10 kV and 130 kV power supplies, C$_1$ and C$_2$ -– correspondingly 1 nF and 250 pF capacitors, D$_1$ and D$_2$ -- distance between the electrodes of 16 cm and 12 cm respectively, L –- lateral displacement of filament with respect to the electrodes.}
\label{schema}
\end{figure}
Three different setups were used in the present work. The filament conductivity measurement, whose results are discussed in Section \ref{sec:conductivity}, relied on the setup 1 described in Fig.\ref{schema} with subscripts “1”. It is based on a 10~kV, 7~mA DC generator (PS$_1$) connected to spherical electrodes (diameter 1.2~cm), which could be placed arbitrarily on either side of the laser beam, at various axial and longitudinal positions. One electrode was grounded (G), while the other (HV), separated by a distance D$_1$ = 16 cm, was set to the potential of the high-voltage generator (PS$_1$) and attached to a 1 nF capacitor (C$_1$). In order to investigate the unloading of the capacitor due to the laser induced filament, the generator (PS$_1$) was disconnected and the voltage on the HV electrode was monitored through a voltage divider probe with a 300 G$\Omega$ impedance and a follower amplifier (TL1169). To avoid constant leakage from the capacitor through the voltage probe, a contact between the HV electrode and the probe was periodically established by an Arduino-based device (A$_1$ in Fig.\ref{schema}). 

The discharge/guiding measurements described in section \ref{sec:discharge} relied on the setup 2 (Fig.\ref{schema} with subscripts “2”). It relied on a 130~kV, 200~{\textmu}A maximum current DC generator (PS$_2$). One electrode was connected to the potential and attached to a 250~pF capacitor (C$_2$), while the ground electrode was at a distance D$_2$~=~12~cm. The potential on the HV electrode was monitored through by a 1:20000 voltage divider of 400~G$\Omega$ impedance. 

The setup 3 is used for the plasma density measurement, to which we will refer as capacitive probe measurement. It consists in the measurement of the positive current swing due to the flow of ionic species between two parallel flat electrodes polarized by a voltage 10~kV (Fig. \ref{plama_measurement}(b)).
It is important to notice that the two first setups have their electrodes along the filament propagation, whereas the capacitive probe measurement have the electrodes above and below the filament.

High peak power ultrashort mid-infrared pulses were delivered by a hybrid OPA/OPCPA system described in detail in \cite{andriukaitis_90_2011}. In brief, the system consists of a femtosecond Yb:CaF$_2$ chirped pulse amplifier (CPA), a three-stage optical parametric amplifier (OPA), a grating/prism (GRISM) stretcher, a picosecond 20~Hz Nd:YAG pump laser, a three-stage OPCPA system, and a grating compressor for mid-infrared pulses. 200~fs pulses from the Yb:CaF$_2$ CPA are used to pump the white light seeded OPA based on type-II KTP crystals and operating at 1460~nm central wavelength. The generated 1460~nm signal pulses are subsequently stretched in a GRISM stretcher and used as a seed for a three-stage OPCPA system based on type II KTA crystals and pumped by 100-ps Nd:YAG laser pulses with energies of respectively, 50, 250, and 700~mJ. The 3.9~{\textmu}m idler pulses are picked up in the 2nd OPCPA-stage and after amplification in the 3rd stage are compressed to 90~fs with a diffraction gratings based compressor. The energy of the compressed 3.9~{\textmu}m pulses exceeds 25~mJ.

Data obtained with the high peak power ultrashort mid-infrared pulses were compared with data acquired by using two different near IR laser systems: namely a 1030~nm Yb:CaF2 CPA generating 220~fs 110~mJ pulses at 50~Hz repetition rate \cite{andriukaitis_220-fs_2015} and a Ti:Sapphire CPA operating at 800~nm central wavelength and delivering 14~mJ 60~fs pulses at a variable repetition rate from 20~Hz to 1~kHz (Coherent Legend).

The experiments with all three laser systems were performed in similar conditions, at a room temperature of 20$^{\circ}$C and relative humidity of 30\%, which corresponds to a background resistivity of air of about 3.10$^{14}\Omega.$m; filamentation in all three cases was assisted by 1 m focusing while electric parameters were kept identical.

The filament side luminescence was recorded with digital photo cameras. The settings of the camera, such as f-number, exposure time and sensitivity, were set in order to avoid saturation of the CMOS-sensor. As it can be seen in Fig.\ref{plama_density}c, the spectral sensitivity of the cameras, used in the experiments, overlaps with the red side of the plasma emission spectrum . 

\section{Results and Discussion}
\subsection{Plasma density along the filament \label{sec:plasma density}}


\begin{figure}[h]
\center\subfigure{\centering\includegraphics[width=0.7\textwidth]{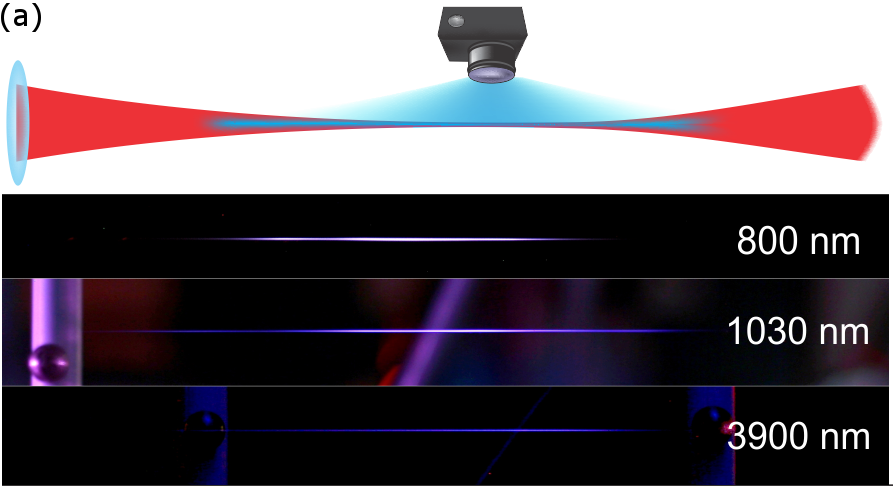}}\\
\subfigure{\centering\includegraphics[width=0.9\textwidth]{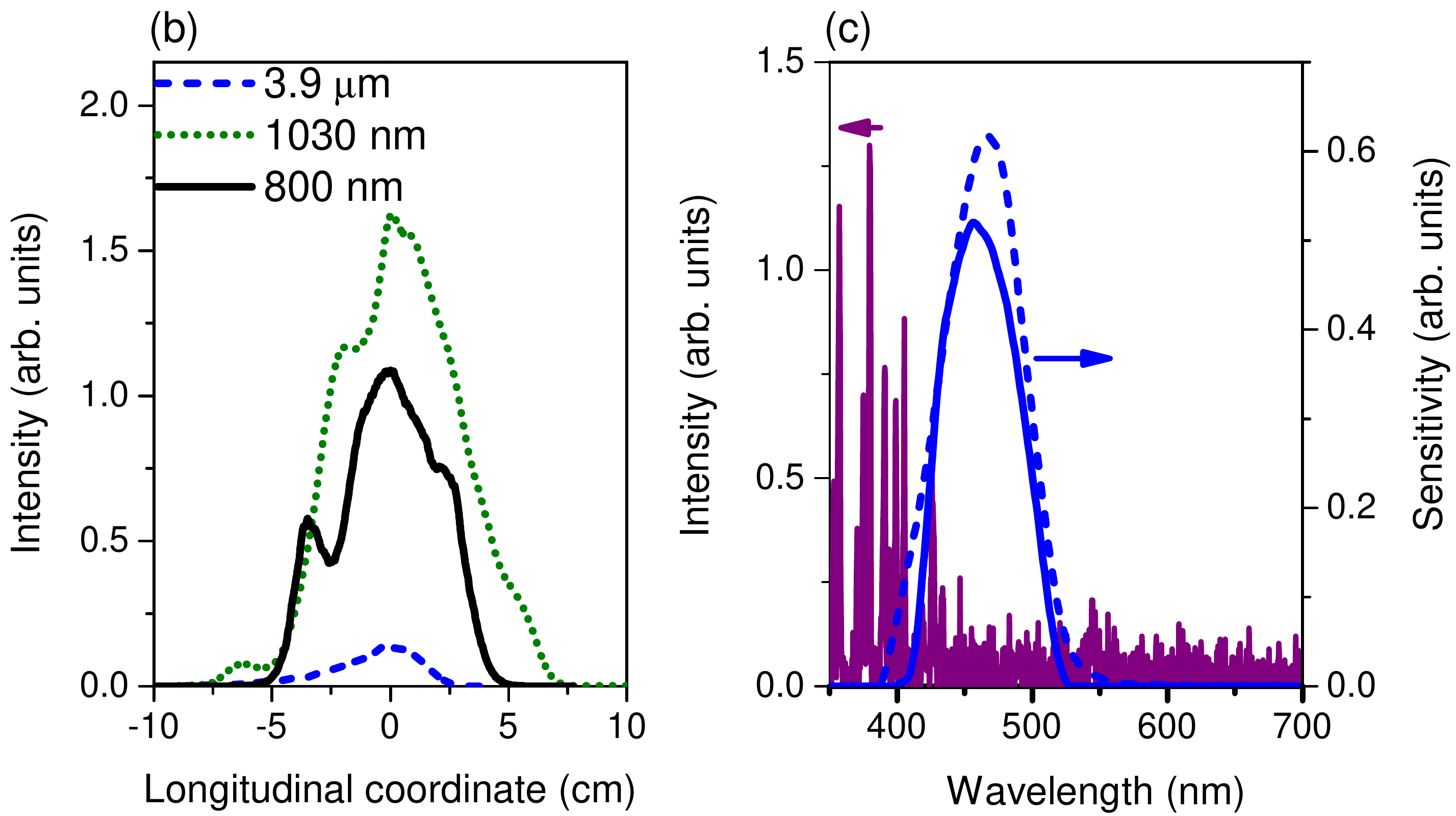}}
\caption{(a) Photographs of the filaments generated at different wavelengths (indicated in the panels) recorded by a digital photo camera; (b) relative amplitudes of extracted spatial luminescence distribution shown in (a). (c) Side-luminescence spectrum of a filament generated by the 3.9~{\textmu}m driver recorded with a spectrometer and spectral sensitivity curves of the blue channels of Canon (400D) (solid line) and Nikon (70D) (dashed line) photo cameras \cite{database}. }
\label{plama_density}
\end{figure}

Plasma emission was recorded at comparable peak powers for all 3 lasers: 14~mJ pulse energy, $\sim$0.23~TW peak power for the 800~nm Ti:Sapphire laser, 36~mJ pulse energy, $\sim$0.16~TW peak power for the 1030~nm Yb:CaF2 laser and 26~mJ pulse energy, $\sim$0.28~TW peak power for the mid IR OPCPA. 

\begin{figure}[h]
\subfigure{\centering\includegraphics[width=0.45\textwidth]{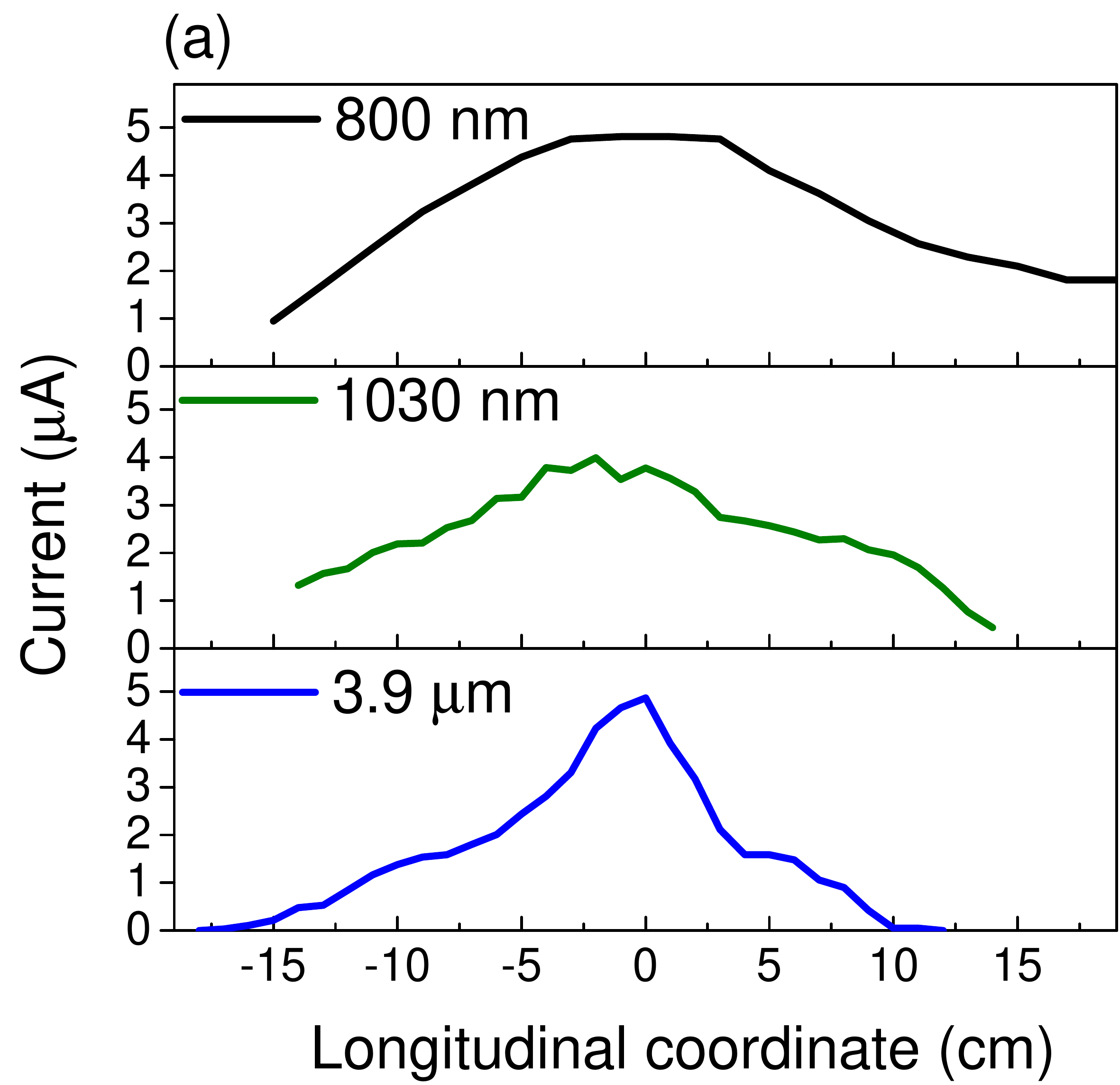}\label{plasma_current_a}}
\subfigure{\centering \raisebox{2cm}{\includegraphics[width=0.45\textwidth]{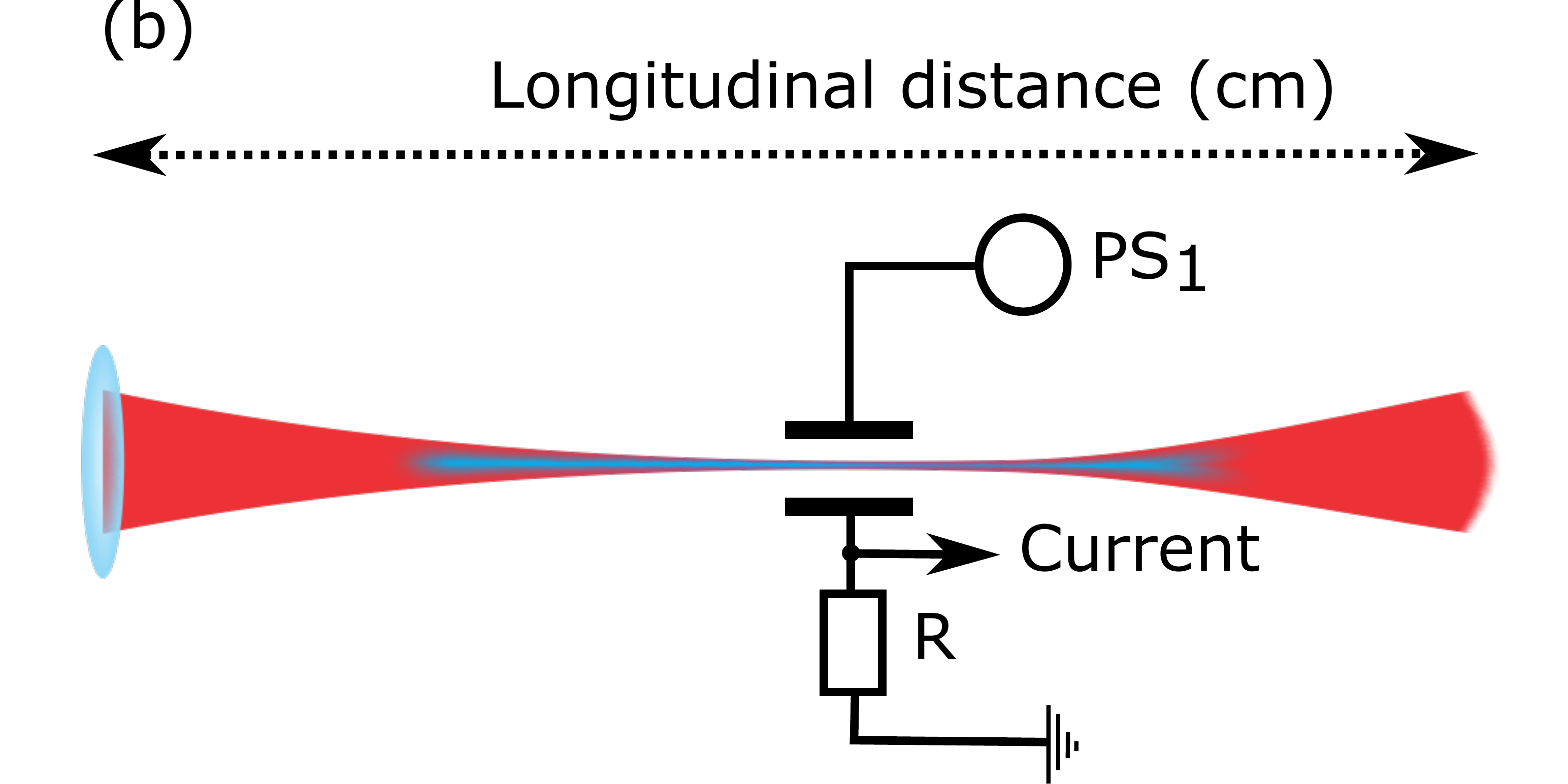}}\label{plasma_current_b}}

\caption{(a) Measured ionic current induced by the filaments at 3 different driving wavelengths (indicated in the panels); (b) schematics of the measurement}
\label{plama_measurement}
\end{figure}

As already reported from numerical simulations for the 3.9~{\textmu}m driver \cite{mitrofanov_mid-infrared_2015,panagiotopoulos_super_2015} and visual observations for the 800~nm driver \cite{polynkin_mobilities_2012} , the visible part of the filaments represents only a fraction of the plasma channel recorded by a transverse electrical probe. Therefore, to assess the plasma density distribution along the filaments more quantitatively, we measured the positive current swing due to the flow of ionic species thanks to the setup 3 described in Fig. \ref{plama_measurement}(b).
The results were analysed according to \cite{polynkin_mobilities_2012} and \cite{henin_contribution_2009}, and provide a good estimate of the total number of positive and negative O$_2$ ions generated in a filament. The use of the dipole induced by the electrons and parent ions just after ionization for the assessment of the initial electron density remains controversial \cite{abdollahpour_measuring_2011}.

The measurements show that the maximum recorded current, and thus the maximum number of ions (positive and negative) per filament length unit (and per pulse) that reach the electrodes is similar for the three wavelengths and correspond roughly to $\sim$10$^9$ elementary charges per cm. However, consistently with the plasma emission presented in Fig.\ref{plama_density}, the length of the plasma column is twice longer at 800~nm than at 3.9~{\textmu}m, leading to a longer conductive path. 

The diameters of the filaments also differ significantly: from 100-200 um in the NIR\cite{berge_ultrashort_2007,couairon_femtosecond_2007} to 600-800~{\textmu}m FWHM at 3.9~{\textmu}m, as estimated from the width of the luminescent plasma line on a side image. It is therefore reasonable to conclude that the maximum ion density is lower by more than one order of magnitude for the mid IR plasma channel as compared to the NIR. However, a direct link between the initial electron densities and ionic densities after travelling several centimeters in hundreds of microseconds is not straightforward \cite{vidal_modeling_2000,shneider_tailoring_2011,papeer_extended_2014}.

Since the peak powers of the laser pulses in all three cases are comparable ($\mathrm{\sim0.16 –- 0.28}$~TW), the number of critical powers reached in the case of mid IR pulses is significantly lower ($\sim$2$-$3~P$_c$(3.9{\textmu}m)) than for the two NIR lasers ($ \mathrm{\sim 20 -– 30} $~P$_c$(NIR)). This might be a cause for the 3.9~{\textmu}m filament being shorter as compared to near-IR filaments: the balance between self-focusing and defocusing is more critical and more sensitive to losses when the laser pulse peak power is close to the threshold of P$_c$. In the case when peak power of mid IR pulses exceeds P$_c$ by more than an order of magnitude a novel regime of filament formation and propagation recently was predicted by numerical modeling \cite{panagiotopoulos_super_2015}.

A final indication of the lower gas ionization in the mid-IR is the total energy loss of the beam after filamentation, ranging from 7\% at 3.9~{\textmu}m to 12\% at 800~nm.

\subsection{Conductivity of mid-IR filaments \label{sec:conductivity}}

In a recently published work, it was shown that 800~nm filaments were able to unload charged capacitors without air breakdown \cite{schubert_remote_2015}. This allows to determine an equivalent resistance $R_{equ}$ of the conductive path between the HV electrode and the ground electrode, from the measured decay time constant $\tau = R_{equ}C_1$, $C_1$ being the capacity attached to the high-voltage electrode, that gets emptied by the current flow. 

We presently investigated the equivalent conductivity of mid-IR filaments thanks to this technique using the setup 1 described in Fig. \ref{schema}, and compare it to the 800~nm and 1030~nm cases at comparable peak powers. The advantage of this method is that it includes all the effects related to the filamentation process, including charge density, charge continuity and space charges, field lines distortion, acceleration by the HV field, shock waves inducing pressure gradients \cite{berge_ultrashort_2007}, etc.

\begin{figure}[h]
\centering\includegraphics[width=0.9\textwidth]{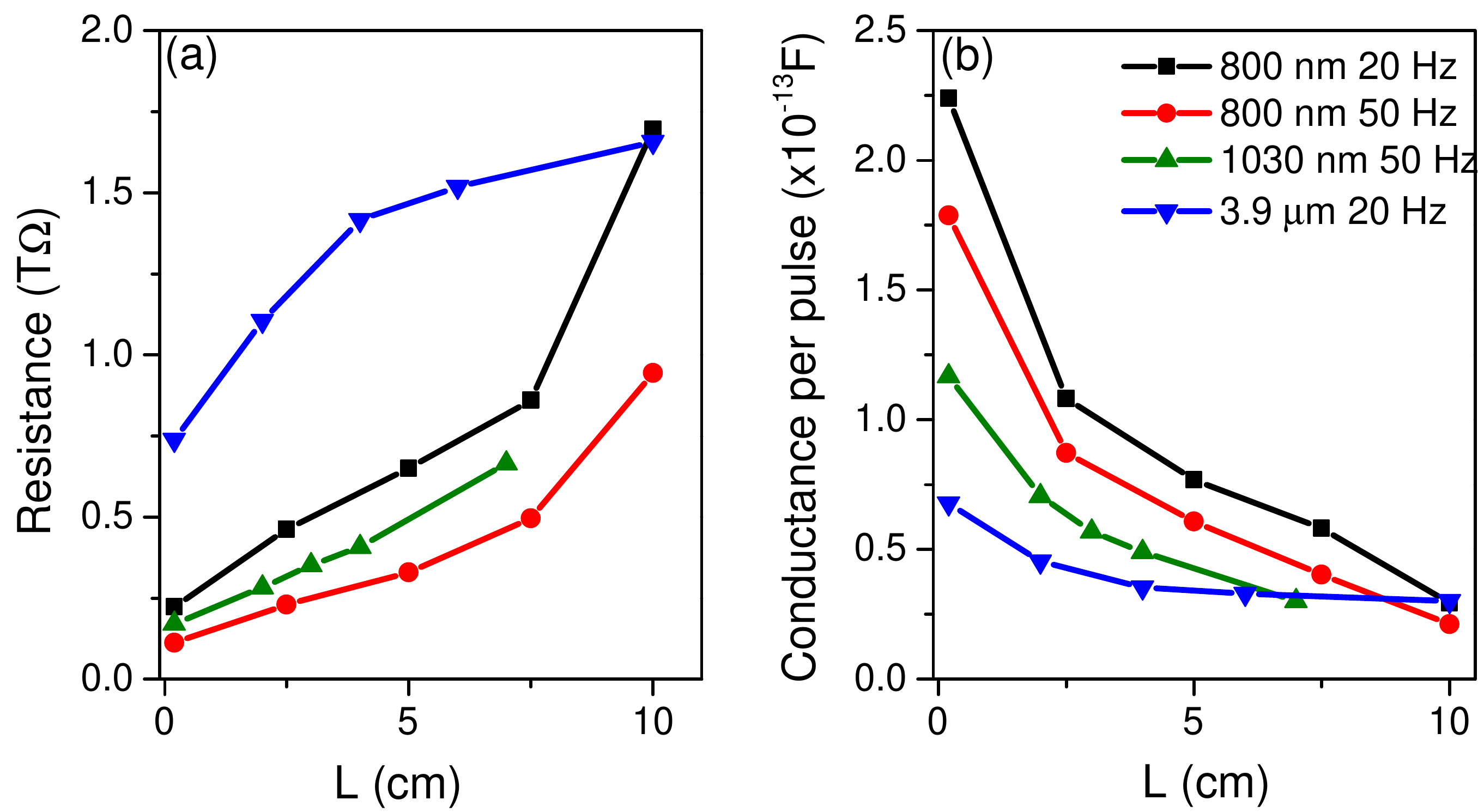}
\caption{(a) Filament-induced capacitor unloading : Equivalent resistance $R_{equ}$ of the path for the 800~nm, 1030~nm and 3.9~{\textmu}m filaments, as a function of the lateral displacement L (see  Fig.\ref{schema}) of the filament relative to the electrodes; (b) normalized filament conductance (see text), as a function of lateral distance L}
\label{unloading} 
\end{figure}

Fig.\ref{unloading}(a) shows the filament induced neutralization of a C$_1$~=~1~nF capacitor loaded at 10~kV, attached to a spherical electrode separated by 16~cm from the ground electrode. The filament propagated parallel to the line connecting the centers of the electrodes with the lateral distance L between the filament and electrodes being varied from 0 to 10~cm (Fig.\ref{schema}).


For the minimal lateral distance (filament grazing to the electrodes) the resistance of the mid-IR filament amounts to 700 $\mathrm{G\Omega}$ over the 16 cm gap, while for the 1030~nm and 800~nm drivers it was 190 $\mathrm{G\Omega}$ and 150 $\mathrm{G\Omega}$ respectively. Here we should note that the repetition rate of the 1030~nm laser was 50~Hz as compared to 20~Hz in the case of 3.9 {\textmu}m OPCPA system. Fig.\ref{unloading}(a) reveals that as the lateral distance increases, the fraction of the filamenting path over the whole path decreases, and the overall equivalent resistance increases accordingly, up to the value of neutral air.

In order to take into account the difference in repetition rates, we normalized the conductance through the repetition rate $\nu$ of the laser. The resulting quantity $C=1/(R_{equ}\nu)$, homogeneous in units of a capacity, could be interpreted as a fraction of the capacity that could be fully extracted, under the considered experimental conditions, by a single laser pulse. 

As displayed in Fig. \ref{unloading}(b), the 800 nm laser filament is the most efficient neutralizer in this configuration, with the capacitance being C(800 nm)~$\sim$~0.2~pF, then C(1030 nm)~$\sim$~0.12~pF and C(3.9 um)~$\sim$~0.07~pF for the filaments grazing on the electrodes. The observed spectral dependence is consistent with the plasma density measurements.
Although each of the three lasers can remotely discharge our capacitor without grazing the electrodes. Over lateral distances of several cm, resistance between the electrodes and the filament is mainly due to air resistivity and each three laser have the same neutralizing capability. We can thus say that 800~nm filament neutralizing capacity have a wider lateral extend than the 1030~nm and 3.9~{\textmu}m ones.

Surprisingly, laser pulses at 800~nm are more efficient to neutralize the charged capacitor when the repetition rate is lower. This counter-intuitive effect will be addressed in future work.


It is interesting to notice that three filaments in our conditions can remove a capacity of the order of 0.1$\sim$pF per shot, which corresponds to the neutralization of typically 6$\times$10$^9$ carriers for the first laser shot with the initial charge in the capacitor corresponding to 6.25$\times$10$^{13}$ carriers. Based on the recorded currents presented in Fig. \ref{plama_measurement}, we estimated the resistance of a pathway through the filament at each wavelength. We assume that the current is proportional to the number of carriers per unit length of filament, and that the resistivity is inversely proportional to this carrier density. The resistance of the filaments at each wavelength can then be compared by integrating the inverse signals shown in Fig.\ref{plama_measurement} along the gap between the electrodes. This estimation yields the difference by a factor of 1.4 between the resistances of 1030~nm and 3.9~$\mathrm\mu$m filaments, and by a factor of 2 between the 800~nm and 3.9~{\textmu}m filament. Those ratios are consistent with the wavelength dependence of the filament conductance (Fig. \ref{unloading}(b)). The difference in conductance between the respective wavelengths is therefore due to the different longitudinal charge distributions over the gap between the electrodes. 

\subsection{Discharge triggering/guiding \label{sec:discharge}}

Charging electrodes separated by 12~cm with 130~kV DC allows investigating a very different action of filaments (setup 2 described in Fig. \ref{schema}). At such high electric field densities, corona effects around the electrodes and impact ionization by electrons are leading to avalanche ionization and heating of the plasma channel, which then become the dominant processes \cite{vidal_modeling_2000,rizk_high_2014}. Breakdown without laser was observed when the distance between electrodes was reduced to 10~cm.

\begin{figure}[h]

\subfigure[]{\centering\includegraphics[height=2.5cm]{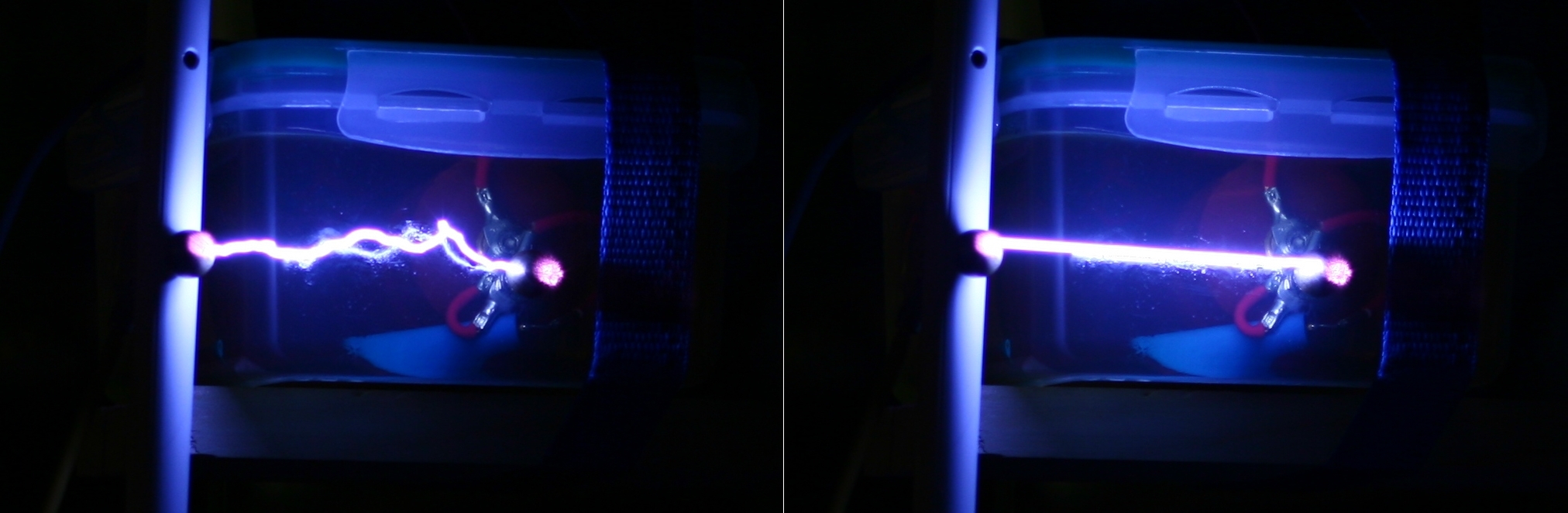} \label{hvdischargea}}
\subfigure[]{\centering\includegraphics[height=2.5cm]{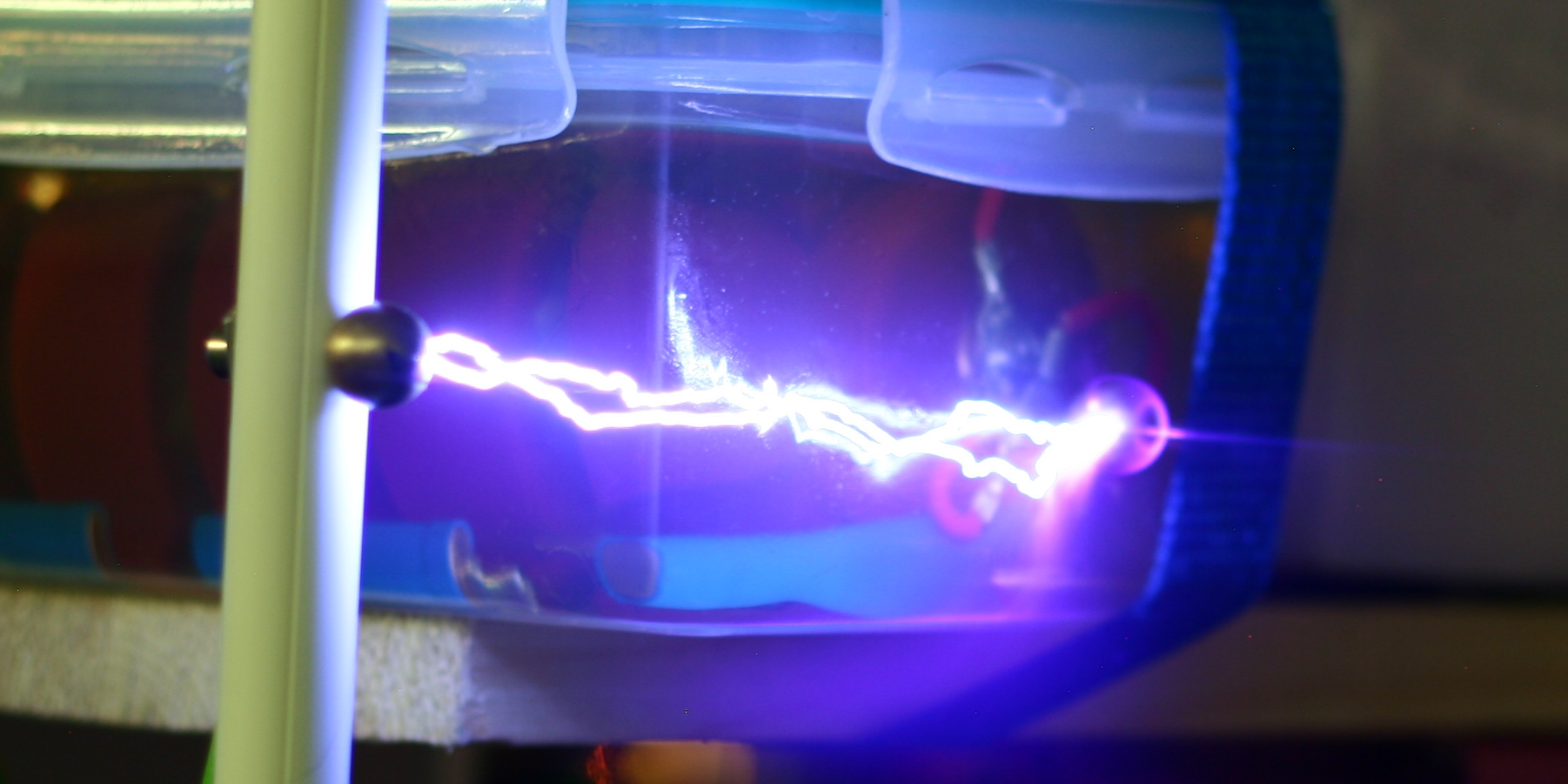} \label{hvdischargeb}}
\caption{(a) Influence of near and mid-IR laser filaments on 130~kV discharges. Triggering and guiding by 1030~nm filament (left: laser beam is blocked; right: laser beam is open); (b) absence of effect in the case of 3.9~{\textmu}m filament}
\label{hvdischarge} 
\end{figure}

A striking difference was observed when the NIR and the MIR filaments were ignited grazing to the charged electrodes (see Fig. \ref{hvdischarge}): as the 1030~nm and 800~nm filaments efficiently triggered (as evidenced by the synchronization of the discharges with the laser shots) with more than 50\% efficiency and guided the HV discharges, neither triggering nor guiding was induced by the mid-IR filament.

Guiding of discharges mainly relies on the creation of a low pressure channel by the heated plasma shockwave \cite{vidal_modeling_2000,tzortzakis_femtosecond_2001,comtois_triggering_2000}. The depth, and accordingly the density gradient, of the low-pressure channel is related to temperature, and thus ultimately to the energy deposited in the filament by the laser. As mentioned in Section \ref{sec:plasma density}, the energy loss measured for the mid IR filament is only 7\% because of lower ionization rates. Furthermore, due to the much larger filament diameter in the mid-IR, this lower energy is deposited in a volume larger by at least one order of magnitude, so that the deposited energy density and the associated heating and air density depletion are 10 to 100 times lower than in the near-IR, which can be expected to be insufficient to guide the discharge. 

\begin{figure}[h]
\centering\includegraphics[width=1\textwidth]{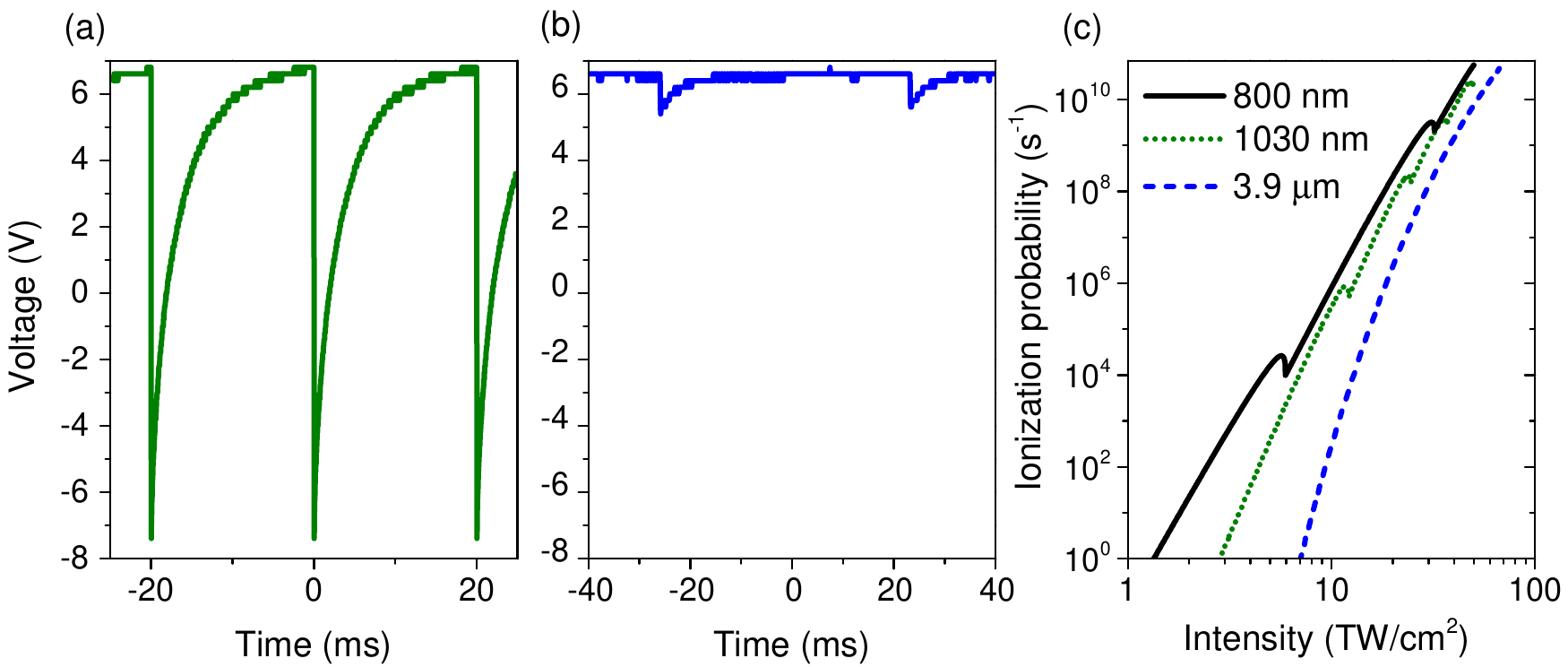}
\caption{Voltage drop measured on electrode charged to 130-kV potential for the (a) 1030~nm and (b) the 3.9~{\textmu}m filaments; (c) ionization rates calculated according to the PPT theory \cite{perelomov_ionization_1966}, for 800~nm, 1030~nm and 3.9~{\textmu}m lasers. }
\label{HV} 
\end{figure}

The different behaviour of NIR and MIR filaments with regard to discharge triggering was quantified by measuring the voltage drop due to the filament induced discharge. As it is shown in Fig. \ref{HV}(a), the NIR Yb:CaF$2$ laser drives a clear voltage drop in our measurement setup, which is a combination of the filament driving carriers away from the electrodes and of induction processes in our setup due to the fast intensity changes. In contrast, the 3.9~{\textmu}m filament induces only a tiny voltage drop, in the order of 5\%. Knowing that avalanche at these high fields is highly dependent on the electron density, this result indicates that the 3.9~{\textmu}m filament, although producing a similar quantity of ions, produces a much lower electron density due to its larger volume. 


This assumption is confirmed by the ionization probability calculated according to the Perelomov-Popov-Terent'ev (PPT) theory \cite{perelomov_ionization_1966}, as shown in Fig. \ref{HV}(c).


The ionization yield, as expected, is widely favored by shorter wavelengths, when the laser intensity lies in the quasi-pure multi-photon regime ($<$ TW/cm$^2$). At higher intensities, however, tunneling ionization dominates and ionization rates converge to similar values. In our case, the filament clamping intensities lie at around 20 TW/cm$^2$ for the 3.9 {\textmu}m \cite{mitrofanov_mid-infrared_2015,panagiotopoulos_super_2015} and around 50-60 TW/cm$^2$ for the near infrared lasers \cite{couairon_femtosecond_2007,berge_ultrashort_2007}. Accordingly,  the near IR lasers release electrons from molecules much more efficiently. Since triggering of HV discharges critically depends on the electron density and temperature \cite{vidal_modeling_2000}, rather than the total content of ions, the mid IR pulses are indeed much less prone, at least at these energies, to generate strong currents, as it was observed experimentally. The lower energy density deposited from the mid IR laser in a larger filament volume confirms these results, with a lower pressure depression due to the weaker heating. Notice, however, that the steepness of the curves in Fig. \ref{HV}(c) precludes more advanced modeling of the discharge, as for instance, a factor 2 error on the clamping intensity of the mid-IR filament would result in an order of magnitude higher ionization rate.

\section{Conclusion}

We compared the electric properties of mid-IR filaments to their more common NIR counterparts in specific, but similar conditions. The main results can be summarized as follows: 
\begin{enumerate}[noitemsep,nolistsep]
\item the total number of ionic carriers created during filamentation ignited with laser pulses at wavelengths of 800~nm, 1030~nm, and 3900~nm is similar (see Section \ref{sec:conductivity} and \ref{sec:plasma density});
\item the initial electronic density is lower for the mid-IR filament (see Section \ref{sec:discharge});
\item the energy deposited in the filament is lower for the mid-IR filament;
\item the above listed properties of mid-IR filaments reduce the ability of triggering and guiding of HV (field strength of $\sim$10kV/cm) discharges.
\end{enumerate}

The results also evidence the importance of the filament length, or more precisely the effective carrier density along the propagation path. For instance, the equivalent resistance measurements emphasize that the total resistance within the path is dominated by the length of the weakly ionized or neutral section, rather than the local absolute carrier density. In real-scale experiments, such as creating a conductive path to thunderstorm clouds, the generation of long, ionized, filaments is crucial. Mid-IR filaments that combine a substantial conductivity and long propagation length \cite{mitrofanov_mid-infrared_2015,panagiotopoulos_super_2015,geints_single_2014} are good candidates for a good trade-off between these two requirements, especially if further laser development allow them to reach the multi TW regime (e.g. several 100 mJ, $<$ 100 fs).

\section*{Acknowledgement}

We acknowledge support from the ERC advanced grant “FILATMO” and the technical support of C. Barreiro (University of Geneva). We also acknowledge the numerical simulations performed by N. Berti (University of Geneva). From the Austrian side the research was supported by the Austrian Science Fund (FWF) through the grants NextLite PO3 (project No.F4903-N23) and MIR Filament (project No.P26658-N27)

%

\end{document}